\documentstyle{article}

\newcommand{\sect}[1]{\section{#1}\indent}
\newcommand{\EQ}{\begin{equation}}
\newcommand{\EN}{\end{equation}}
\newcommand{\bea}{\begin{eqnarray}}
\newcommand{\ena}{\end{eqnarray}}
\newcommand{\vs}[1]{\vspace{#1 mm}}

\renewcommand{\b}{\beta}
\renewcommand{\c}{\gamma}

\newcommand{\shalf}{\frac{1}{2}}
\newcommand{\pa}{\partial}

\newcommand{\nn}{\nonumber \\}

\begin{document}
\topmargin 0pt
\oddsidemargin 5mm

\renewcommand{\Im}{{\rm Im}\,}
\newcommand{\NP}[1]{Nucl.\ Phys.\ {\bf #1}}
\newcommand{\PL}[1]{Phys.\ Lett.\ {\bf #1}}
\newcommand{\CMP}[1]{Comm.\ Math.\ Phys.\ {\bf #1}}
\newcommand{\PR}[1]{Phys.\ Rev.\ {\bf #1}}
\newcommand{\PRL}[1]{Phys.\ Rev.\ Lett.\ {\bf #1}}
\newcommand{\PTP}[1]{Prog.\ Theor.\ Phys.\ {\bf #1}}
\newcommand{\PTPS}[1]{Prog.\ Theor.\ Phys.\ Suppl.\ {\bf #1}}
\newcommand{\MPL}[1]{Mod.\ Phys.\ Lett.\ {\bf #1}}
\newcommand{\IJMP}[1]{Int.\ J.\ Mod.\ Phys.\ {\bf #1}}

\begin{titlepage}
\setcounter{page}{0}
\begin{flushright}
NBI-HE-94-51, OU-HET 203 \\
November 1994\\
hep-th/9411156
\end{flushright}

\vs{8}
\begin{center}
{\Large NOTE ON $W_3$ REALIZATIONS OF THE BOSONIC STRING}

\vs{20}
{\large Fiorenzo Bastianelli\footnote{e-mail address:
 fiorenzo@nbivax.nbi.dk}}\\
{\em The Niels Bohr Institute, Blegdamsvej 17, DK-2100 Copenhagen \O,
Denmark}\\

\vs{8}
{\large Nobuyoshi Ohta\footnote{e-mail address:
ohta@phys.wani.osaka-u.ac.jp}}\\
{\em Department of Physics, Osaka University, Toyonaka, Osaka 560, Japan}
\end{center}

\vs{30}
\centerline{{\bf{Abstract}}}

In order to investigate to what extent string theories are different
vacua of a general string theory (the ``universal string"),
we discuss realizations of the bosonic string as particular
background of certain types of $W$-strings.
Our discussions include linearized $W_3^{lin}$, non-critical $W_3$,
linearized $W_3^{(2)lin}$ and critical $W_3^{(2)}$ realizations of
the bosonic string.

\end{titlepage}
\newpage
\renewcommand{\thefootnote}{\arabic{footnote}}
\setcounter{footnote}{0}

\sect{Introduction}
\indent
Along the line of realization of various string theories in
those with higher worldsheet gauge symmetries~\cite{R},
those based on nonlinear algebras seem to
belong to a very special class.
The nonlinear character of these algebras makes it difficult to
construct embeddings, and up to now only few special
realizations have been discovered for the case of the bosonic
string embedded in the $W$-strings~\cite{W,KST,BBR}. It would be quite
interesting to examine further such realizations
with the aim of finding general methods to describe
embeddings. Our motivations arise from investigating to what
extent string theories can be viewed as
different vacua of a general string theory
(the ``universal string"). In this note we will present a few
remarks and a few additional realizations
of the bosonic string as particular background of certain types of
$W$-strings. Our main idea is to use linearized versions of the
$W$-algebras to find embeddings and to map these embeddings back to
the nonlinear basis of the algebra. We could
find embeddings for the linearized version of the $W$-algebras
and prove their equivalence to the bosonic string, but we encountered
technical difficulties in proving the equivalence in the
nonlinear cases.

\sect{$W_3^{lin}$ realization of the bosonic string}
\indent
An interesting approach to realize embeddings of the various string
theories into those based on nonlinear algebras is to consider the
linearization of these nonlinear
algebras by including additional symmetry generators
and performing nonlinear redefinitions of the generators.
Recently Krivonos and Sorin~\cite{KS} have constructed a linearized version
of the $W_3$ algebra, which is denoted by $W_3^{lin}$ and is given by
\bea
T(z) T(w) &\sim&  \frac{c}{2} \frac{1}{(z-w)^4} + \frac{2 T(w)}{(z-w)^2} +
\frac{\partial T(w)}{(z-w)} \nn
T(z) G(w) &\sim& \frac{x_1 G(w)}{(z-w)^2}+\frac{\partial G(w)}{(z-w)} \nn
T(z) J(w) &\sim& \frac{J(w)}{(z-w)^2} +\frac{\partial J(w)}{(z-w)} \nn
J(z) J(w) &\sim& \frac{x_2}{(z-w)^2} \nn
J(z) G(w) &\sim& \frac{G(w)}{(z-w)} ,
\label{w3lin}
\ena
where the central charge $c$ and the coefficients $x_1,x_2$
are parametrized by
\EQ
 c = \frac{1 - 4 x - 9 x^2 }{1 +x} ,\ \  x_1 = \frac{3}{2} +
\frac{1}{1+x} ,\ \  x_2 = 1+x .
\EN
An invertible nonlinear transformation maps this algebra into
the usual nonlinear $W_3$ algebra extended by a spin $1$ current.
Such an extended $W_3$ algebra, which we will call $W_{3,1}$,
is unique if one
requires to preserve the usual $W_3$ algebra as subalgebra.
This was discussed in ref.~\cite{KS}, to which
we refer for the complete OPE of the $W_{3,1}$ algebra.
In this section we will show how the bosonic string can be realized
as a special case of the more general string that can be
constructed by gauging the $W_3^{lin}$ algebra.

First we construct the BRST charge for the $W_3^{lin}$ algebra.
It is given by
\EQ
Q = \oint [
c_t T + c_g G + c_j J + c_t ( \partial c_t b_t  + \partial b_g c_g
- b_j \partial c_j ) + x_1 \partial c_t b_g c_g + c_j b_g c_g ],
\label{brst}
\EN
with  the  ghost correlators given by
\EQ
c_\alpha(z) b_\beta(w) \sim \frac{\delta_{\alpha \beta}}{(z-w)} ,
\ \ \ \ \ \ \ \
\alpha,\beta = t,g,j.
\EN
It is nilpotent for $x=-2$. This value implies
$ c= 27,\ x_1 = \frac{1}{2},\  x_2 = -1 $.
Note that the current $G$ acquires spin $\shalf$ at the critical value
$x=-2$, but it still has even Grassman parity.

Now we show that the bosonic string can be embedded into
the more general string based on the $W_3^{lin}$ algebra.
Consider a consistent bosonic string background described by a
conformal field theory
with stress tensor $T_m$ satisfying the Virasoro algebra with central
charge $c=26$. Tensoring it with a set of two commuting $bc$ systems,
$(\b_g,\c_g)$, $(\b_j,\c_j)$ of spin $(\shalf,\shalf)$, $(1,0)$,
respectively, one can construct the following realization
of the critical $W_3^{lin}$ algebra
\bea
T &=& T_m + T(\b_g,\c_g) + T(\b_j,\c_j),  \nn
G &=& \b_g , \nn
J &=& \b_j + \b_g \c_g ,
\label{rea}
\ena
where we denote by $T(b,c)$ the stress tensor of a $(b,c)$ system
of spin $(\lambda, 1 - \lambda)$
\EQ
T(b,c)= - \lambda b \partial c + (1- \lambda) \partial b c \ .
\EN
Plugging the particular realization (\ref{rea}) into eq.~(\ref{brst}),
we get the following nilpotent BRST charge
\bea
\tilde Q &=& \oint[
c_t ( T_m + \frac{1}{2} T(b_t,c_t) + T(b_g,c_g) + T(b_j,c_j) +
T(\b_g,\c_g) + T(\b_j,\c_j) ) \nn
& & + c_g \b_g  + c_j ( \b_j + \b_g \c_g + b_g c_g) ].
\label{brst1}
\ena
It is canonically equivalent to the BRST charge of the bosonic string
plus a sector of non-minimal terms which decouples
and has trivial cohomology. In fact one can check that
\EQ
{\rm e}^{R_2} {\rm e}^{R_1} \tilde Q {\rm e}^{-R_1} {\rm e}^{-R_2}
= Q_{bos} + Q_{nm} ,
\EN
where $Q_{bos}$ is the BRST of the usual bosonic string
\EQ
Q_{bos} = \oint c_t \left( T_m + \frac{1}{2} T(b_t,c_t) \right) ,
\label{bosbrst}
\EN
and where
\bea
Q_{nm} &=& \oint ( c_g \b_g  + c_j  \b_j ), \nn
R_1 &=& - \oint  c_j b_g \c_g , \nn
R_2 &=& - \oint c_t \left( T \left(b_g,\c_g\right) +
T \left(b_j,\c_j\right) \right).
\label{xx}
\ena
The topological charge $Q_{nm}$ imposes the
constraint that the fields $(\b_g,\c_g,b_g,c_g)$ as well as
$(\b_j,\c_j,b_j,c_j)$ make quartets and decouple from the theory.
Thus the bosonic string is realized as a particular background
of the $W_3^{lin}$-string.

\sect{Finding quadratic $W_3$}
\indent
The realization (\ref{rea}) can be used to find a realization of
nonlinear $W_3$ as follows. Define the generators
\bea
T_W &=& T - \frac{5}{2}\pa J \nn
&=& T_m - 3 \b_g\pa\c_g - 2 \pa\b_g\c_g - \b_j\pa\c_j -
\frac{5}{2} \pa\b_j,  \nn
W &=& G + 2i \sqrt{\frac{2}{133}} \left[ J T + \frac{2}{3} J^3
 + \frac{5}{2} J\pa J + \frac{5}{4} \pa T + \frac{25}{24} \pa^2 J\right]
\nn
&=& \b_g + i \sqrt{\frac{2}{133}}\left( 2 \b_g \c_g T_m + 2 \b_j T_m
+ \frac{4}{3} \b_j^3 + 4 \b_g\c_g\b_j^2 + 5 \b_g\c_g\pa\b_j
+ 4 \b_g^2\c_g^2\b_j \right.
\nn &+& \frac{4}{3}\b_g^3\c_g^3
+ 10 \b_g\pa\b_g\c_g^2 + 10 \pa\b_g\c_g\b_j
- 2 \b_g\c_g\b_j\pa\c_j + 7 \pa^2\b_g\c_g
- \shalf \pa\b_g\pa\c_g
\nn  &-& \left. \shalf \b_g\pa^2\c_g
+ 5 \b_j\pa\b_j - 2 \b_j^2\pa\c_j - \frac{5}{2}\pa\b_j\pa\c_j
- \frac{5}{2}\b_j\pa^2\c_j + \frac{5}{2}\pa T_m
+ \frac{25}{12}\pa^2\b_j \right),
\label{realw}
\ena
where $T,G,J$ are given in (\ref{rea}).
One may verify that the generators $(T_W, W,J)$
realize a representation of the $W_{3,1}$ algebra with the
correct amount of central charge ($c=102$)
expected to balance the ghost
contribution. Thus one may conjecture that
it defines a critical realization of the $W_{3,1}$ algebra.
However to prove such a conjecture one has to construct
the BRST charge for $W_{3,1}$ and verify that the above realization
makes it nilpotent. The construction of such a BRST charge
is tedious because the algebra closes with cubic relations and it
is not discussed here.

An intriguing alternative is to notice that $(T_W, W)$ in (\ref{realw})
satisfy the $W_3$ OPE with central charge $c=102$.
This realization is interesting in that the generator
$W$ contains a linear term $\b_g$, indicating the spontaneous breakdown
of the symmetry. This is quite different from the realizations discussed
in refs.~\cite{W,BBR}, in which there is a linear term but with
derivatives.
However the algebra satisfied by these generators has the non-critical
central charge $c =102$ and we have the ``extra" matter $\b_j,\c_j$.
One can still define a nilpotent BRST operator by coupling the $W_3$
matter system described by $(T_W,W)$
to a ``Liouville'' system \`{a} la ref.~\cite{NON}.
For this purpose, we need a $W_3$ realization with central charge
$c=-2$ as a ``Liouville" system.
We find that
\bea
T_L &=& -\eta_L\pa\xi_L, \nn
W_L &=& \frac{1}{\sqrt{6}}( \pa\eta_L\pa\xi_L - \eta_L\pa^2\xi_L),
\ena
satisfy the $W_3$ algebra with $c=-2$, where $(\eta_L,\xi_L)$ are
anticommuting fields.
Using this realization, the nilpotent BRST operator is given by
\bea
Q &=& \oint \left[ c_t (T_W + T_L) + i \frac{\sqrt{133}}{2} c_w W
 + \frac{\sqrt{3}}{2} c_w W_L + (T_L-T_W) b_t c_w \pa c_w \right.\nn
&& \left. + c_t\pa c_t b_t - 3 c_t b_w \pa c_w -2 c_t \pa b_w c_w
 - \frac{13}{8} \pa b_t c_w \pa^2 c_w -\frac{65}{24} b_t c_w \pa^3 c_w
\right].
\label{3.3}
\ena
Since we have the linear term $\b_g$ in $W$ which can be used to decouple
$(\b_g,\c_g,b_w,c_w)$ as in the superstrings~\cite{R}, we expect that
this model contains the bosonic string as a subsector without states
from these fields. Unfortunately a constraint that would eliminate the
fields $(\eta_L,\xi_L,\beta_j, \gamma_j)$ does not seem to be
present in the BRST charge, so it may be that additional states are
present, as in the model of ref.~\cite{BBR}  where a construction
reminiscent of ours has been presented. It would be interesting
to examine the full cohomology of eq.~(\ref{3.3}).

Before closing this section, we would like to note that
eq.~(\ref{realw}) suggests that it may be possible to find a
critical realization using only $(\b_g,\c_g)$ with a linear $\b_g$
term in $W$, unlike that of refs.~\cite{W,BBR}.
However terms like $\pa^2\gamma_g$ having zero dimension
can appear with an arbitrary power, making it harder to proceed
systematically in the construction of the generators.
For example, a possible method to find such a realization is to use
an expansion in ghost number as in ref.~\cite{BO}. Namely we assign
ghost number $-1$ to $\b_g$ and $1$ to $\c_g$, and assume the following
form
\bea
T &=& T_m - 3 \b_g \pa \c_g - 2 \pa \b_g \c_g + ({\rm possible \; terms
\; with \; ghost \; number \; 2}) + \cdots, \nn
W &=& \b_g + ({\rm possible \; terms \; with \; ghost \;
number \; 1}) +\cdots.
\ena
Then one can try to satisfy the OPE for $W_3$ at each ghost number.
Unfortunately the calculation involves increasingly large number of
terms for increasing ghost numbers and it seems
difficult to construct a realization in this way.

\sect{$W^{(2)lin}_3$ realization of the bosonic string}
\indent
In ref.~\cite{KS} another nonlinear algebra
$W^{(2)}_3$~\cite{B} was also linearized. The linearized version,
called $W^{(2)lin}_3$, is given by
\bea
T(z) T(w) &\sim& \frac{(7-9 x)x}{2(x+1)}\frac{1}{(z-w)^4}
 + \frac{2 T(w)}{(z-w)^2} + \frac{\pa T(w)}{(z-w)}, \nn
T(z) J(w) &\sim& \frac{J(w)}{(z-w)^2} + \frac{\pa J(w)}{(z-w)}, \nn
T(z) G^\pm(w) &\sim& \frac{3}{2} \frac{G^\pm(w)}{(z-w)^2}
 + \frac{\pa G^\pm(w)}{(z-w)}, \nn
J(z) J(w) &\sim& \frac{x}{(z-w)^2}, \qquad
J(z) G^\pm(w) \sim \pm \frac{G^\pm(w)}{(z-w)}, \nn
T(z) K(w) &\sim& -\shalf\frac{K(w)}{(z-w)^2} + \frac{\pa K(w)}{(z-w)},
\qquad
J(z) K(w) \sim \frac{K(w)}{(z-w)}, \nn
G^-(z) K(w)&\sim& \frac{1}{(z-w)}.
\label{w23}
\ena
The standard procedure for constructing the BRST operator fails
to give a nilpotent BRST charge.
This is presumably related to the fact that not all possible central
charges compatible with the Jacobi identities are present in the
OPE (\ref{w23}). In fact one can write down all possible central charges
$(c,c_1,c_2,c_3,x,x_1)$
that can appear by dimensional analysis
\bea
T(z) T(w) &\sim& \frac{c}{2}\frac{1}{(z-w)^4}
 + \frac{2 T(w)}{(z-w)^2} + \frac{\pa T(w)}{(z-w)}, \nn
T(z) J(w) &\sim& \frac{c_1}{(z-w)^3} +
\frac{J(w)}{(z-w)^2} + \frac{\pa J(w)}{(z-w)}, \nn
T(z) G^\pm(w) &\sim& \frac{3}{2} \frac{G^\pm(w)}{(z-w)^2}
 + \frac{\pa G^\pm(w)}{(z-w)}, \nn
J(z) J(w) &\sim& \frac{x}{(z-w)^2}, \qquad
J(z) G^\pm(w) \sim \pm \frac{G^\pm(w)}{(z-w)}, \nn
T(z) K(w) &\sim& -\shalf\frac{K(w)}{(z-w)^2} + \frac{\pa K(w)}{(z-w)},
\qquad
J(z) K(w) \sim \frac{K(w)}{(z-w)}, \nn
G^-(z) K(w)&\sim& \frac{x_1}{(z-w)}, \qquad
G^+(z) K(w)\sim \frac{c_3}{(z-w)}, \nn
G^+(z) G^-(w)&\sim& \frac{c_2}{(z-w)^3}.
\label{4.2}
\ena
Then the standard BRST charge
\bea
Q &=& \oint \left[
 c_t T + c_j J + c_+ G^+ + c_- G^- + c_k K + c_t \pa c_t b_t
 + c_t \pa b^+ c_+ + \frac{3}{2} \pa c_t b^+ c_+
 + c_t \pa b^- c_-  \right. \nn
&& \left. + \frac{3}{2} \pa c_t b^- c_- - c_t b_j \pa c_j +
c_j (b^+ c_+ - b^- c_-) + c_t \pa b_k c_k -
\shalf \pa c_t b_k c_k + c_j b_k c_k
\right],
\label{4.3}
\ena
is nilpotent for $c=61$, $c_1 =2 $, $x=-3$ and $x_1 =c_2=c_3=0$.
In general these central charges are not all independent and
should be fixed by the Jacobi identities.
Since we are only interested in the critical algebra, we have
not worked out all the relations arising from Jacobi identities.
It is enough to know that the nilpotency of the BRST charge guarantees
that the Jacobi identities are satisfied for the critical algebra,
for which we will indeed find an explicit realization.

The bosonic string can be embedded in the string described by the
algebra (\ref{4.2}) by introducing,
along with a consistent bosonic string background
given by the stress tensor $T_m$, four commuting $bc$ systems
$(\beta^+,\gamma_+)$,  $(\beta^-,\gamma_-)$,
$(\beta_k,\gamma_k)$, $(\beta_j,\gamma_j)$
of spin $(\frac{3}{2},-\frac{1}{2})$, $(\frac{3}{2},-\frac{1}{2})$
$(-\frac{1}{2},\frac{3}{2})$, $(1,0)$, respectively.
The generators
\bea
T &=& T_m + T(\beta^+,\gamma_+) + T(\beta^-,\gamma_-) +
T(\beta_k,\gamma_k) + T(\beta_j,\gamma_j), \nn
G^+ &=& \b^+, \qquad
G^- = \b^-, \qquad
K = \beta_k, \nn
J &=& \b_j + \b^+\c_+ - \b^-\c_- +  \beta_k \gamma_k ,
\label{rea3}
\ena
realize the critical algebra (\ref{4.2}).
Plugging this realization in eq.~(\ref{4.3}), we obtain a BRST
charge $\tilde Q$ that is canonically equivalent to the BRST
charge of the bosonic string $Q_{bos}$
plus a sector of non-minimal terms which decouples
and carries trivial cohomology. In fact one can check that
\EQ
{\rm e}^{R_2} {\rm e}^{R_1} \tilde Q {\rm e}^{-R_1} {\rm e}^{-R_2}
= Q_{bos} + Q_{nm},
\EN
where
\bea
Q_{nm} &=& \oint ( c_+ \b^+ + c_- \b^- + c_k \b_k  + c_j  \b_j ), \nn
R_1 &=& - \oint  c_j ( b^+ \c_+ - b^- \c_- + b_k \c_k) , \nn
R_2 &=& - \oint c_t \left( T(b^+,\c_+) +T(b^-,\c_-) + T(b_k,\c_k) +
T(b_j,\c_j) \right).
\ena

One problem with this construction
is that the nilpotency of the BRST charge
(\ref{4.3}) requires $x_1=0$, but  this must be non-vanishing in
order to reproduce the $W_3^{(2)}$ algebra as a subalgebra of (\ref{4.2}).
Another approach that allows to construct a nilpotent
operator from the algebra (\ref{w23}) and avoids this problem is that of
introducing an additional commuting $(\b,\c)$ system of spin
$(\frac{3}{2},-\shalf)$ so that
\bea
Q &=& \oint \left[
 c_t T + c_j J + c_+ G^+ + c_- G^- + c_k K + c_t \pa c_t b_t
 + c_t \pa b^+ c_+ + \frac{3}{2} \pa c_t b^+ c_+ + c_t \pa b^- c_-
 \right. \nn
&& + \frac{3}{2} \pa c_t b^- c_- - c_t b_j \pa c_j + c_j (b^+ c_+ -
 b^- c_-) + c_t \pa b_k c_k - \shalf \pa c_t b_k c_k + c_j b_k c_k \nn
&& \left. - \frac{3}{2}c_t \b\pa\c - \shalf c_t \pa\b\c
 - c_j \b\c + c_-\b + c_k\c \right],
\label{brst2}
\ena
is nilpotent for $x=-2$ or central charge $c=50$.
Note that the first two lines of eq.~(\ref{brst2}) give the charge
naively expected which however is not nilpotent.
Actually this approach is just a reinterpretation of the previous one.
We can read off the total matter
generators by taking anticommutators of (\ref{brst2})
with the respective ghosts and dropping the ghost part.
We find
\bea
T' &=& T + T(\b,\c), \qquad
J'= J - \b\c, \nn
{G^+}' &=& G^+, \qquad
{G^-}'= G^-+\b, \qquad
K' = K+\c.
\ena
We see that these are modified by $\b,\c$ such that they satisfy
the critical algebra (\ref{4.2}). This suggests
that the generators $G^-$ and $K$ should be spontaneously broken
because the total generators ${G^-}'$ and $K'$ contain linear
terms. It is also possible to eliminate $c_-,b^-,\b,\c$, at
the cost of redefining the generators. The BRST
charge (\ref{brst2}) can be transformed as
\bea
e^{R} Q e^{-R} &=& \oint \left[
 c_t T + c_j J + c_+ G^+ + c_t \pa c_t b_t + c_t \pa b^+ c_+
 + \frac{3}{2} \pa c_t b^+ c_+ - c_t b_j \pa c_j \right.\nn
&& \left. + c_t \pa G^- K + \frac{3}{2} \pa c_t G^- K +
 c_j (b^+ c_+ - G^- K) + c_k K + c_-\b \right],
\label{brst3}
\ena
where
\EQ
R = \oint \left[
- \c G^- - c_t \left( T(b^-,\c) + T(b_k,G^-) \right)
+ c_j (b^- \c- b_k G^-) \right].
\EN
The last term in eq.~(\ref{brst3}) implies that the four fields
$(b^-,c_-,\b,\c)$ make a quartet and decouple from the theory.
The generators corresponding to $G^-,K$ obtained from (\ref{brst3})
by the anticommutators with $b^-,b_k$ decouple from the rest of
the generators. This is an implementation through similarity
transformations of the procedure of eliminating $G^-,K$
by redefinitions given in ref.~\cite{KS}.

We can find a realization of the algebra (\ref{w23}) for
$x=-2$, which is given by
\bea
T &=& T_m + T(\b^+,\c_+ ) + T(\b^-,\c_- ) + T(\b_j, \c_j ), \nn
G^+ &=& \b^+, \qquad
G^- = \b^-, \qquad
K = - \c_- ,\nn
J &=& \b_j + \b^+\c_+ - \b^-\c_- .
\label{rea2}
\ena
Now we can show the equivalence to bosonic string. The BRST
charge (\ref{brst2}) with eq.~(\ref{rea2}) substituted in
can be transformed as
\EQ
e^{R_2} e^{R_1} Q e^{-R_1} e^{-R_2} = Q_{bos} + Q_{nm},
\EN
where $Q_{bos}$ is that for the
bosonic string given in eq.~(\ref{bosbrst})
and $Q_{nm}$ is that for the non-minimal sector
\EQ
Q_{nm} = \oint [ c_j \b_j + c_+ \b^+ + c_- (\b^-+\b) - c_k (\c_- - \c)],
\EN
and where
\bea
R_1 &=& \oint \left[
\shalf \c_j ( - \b^+\c_+ + \b^-\c_- + \b\c - b^+c_+ +b^-c_--b_k c_k)
\right. \nn
&& \left. +\frac{1}{4} c_j \left( -2 b^+\c_+ + b^-(\c_- + \c )
- b_k (\b^- - \b) \right) \right], \nn
R_2 &=& -\oint c_t \left[ T(b^+,\c_+ ) + T \left( b^-,\frac{\c_- +\c}{2}
 \right) + T \left( b_k, \frac{\b^- - \b}{2}\right) +T(b_j,\c_j)\right].
\ena
Again the non-minimal term imposes the condition that all fields
except those in the original bosonic string decouple from the physical
subspace and the theory is equivalent to the bosonic string.

\sect{$W^{(2)}_3$ realization of the bosonic string}
\indent
The unorthodox way of constructing a nilpotent BRST charge
for the linear algebra in eq.~(\ref{w23}),
presented in the second part of the previous paragraph,
gives us a hint on how to obtain a realization of the
bosonic string as a particular background for
the nonlinear $W^{(2)}_3$-string.
To show this, we first note that the $W^{(2)}_3 $ BRST
charge~\cite{Khvi} is given by
\bea
Q &=& \oint [
 c_t T + c_+ \hat G^+ + c_- G^- + c_j J  + c_t ( \shalf T(b_t,c_t) +
 T(b^+,c_+) + T(b^-,c_-) + T(b_j,c_j) )
\nn
&+& b_t c_+ c_- + \frac{3}{2} b_j (c_+ \pa c_- - \pa c_+ c_- ) +
c_j (b^+ c_+ - b^- c_-) - \frac{2}{x+1} b_j c_+ c_- J ],
\ena
where the spin $\frac{3}{2}$ generator
$\hat G^+$ is given by the nonlinear redefinition of
Krivonos and Sorin~\cite{KS}
\bea
\hat G^+ &=& G^+ + T K - \frac{2}{x+1} J^2 K - \frac{3 x+7}{2(x+1)}
 \pa J K + \frac{2}{x+1} J G^- K^2 - \frac{2}{3(x+1)} G^- G^- K^3 \nn
 &-& 3 K \pa K G^- + \frac{1-x}{1+x}
 K^2 \pa G^- + 3 \pa (J K) - \frac{3(x+1)}{2} \pa^2 K ,
\ena
such that $(T,\hat G^+, G^-, J)$  constitute a $W^{(2)}_3 $ algebra.
We find that the nilpotency condition for this charge is precisely
$x=-2$ or $c=50$. Hence substituting the critical realization
given in eq.~(\ref{rea2}), we get an embedding of the bosonic string.
The conjecture that this BRST charge is canonically
equivalent to that of the bosonic string plus a sector of non-minimal
fields may be proved by performing similarity transformations, but we
have not been able to find the set of transformations that does the job.

\sect{Conclusions}
\indent
We have presented a few realizations of the bosonic string
as a background for some $W$-strings and discussed their properties.
We have used linearized versions of the $W$-algebras
to identify the embeddings.
For the $W_3$ case we have obtained a non-critical
realization. Adding to it a ``Liouville" system to obtain
criticality we have achieved a critical embedding.
However the resulting
theory presumably contains additional states on top of those
present in the bosonic string. In fact a full equivalence
could not be established. In the $W_3^{(2)}$ case
the linearized embedding has helped us to identify a consistent
background for the $W_3^{(2)}$-string that might reproduce the
full structure of the bosonic string without additional states.
Thus the $W_3^{(2)}$ case seems to be simpler than the $W_3$ case
in that we could identify a critical embedding.
However we have not found the correct similarity transformation
that proves the complete equivalence, even though we suspect
this should be possible.
The $W_3$ and $W_3^{(2)}$ algebras are closely related.
In fact $W$-algebras can generically be derived as hamiltonian
reduction of WZNW models, and the specific case of the $W_3$ and
$W_3^{(2)}$ algebras correspond to different embedding
of $sl(2)$ into $sl(3)$~\cite{BTD}.
Perhaps the method of hamiltonian reduction could be
successfully used in the search for a ``universal string''.

\vs{5}
\noindent
{\it Acknowledgements}

We have checked many of our OPEs using the Mathematica package
developed by K. Thielemans \cite{KT}.
We would like to thank H. Kunitomo and J.L. Petersen for valuable
discussions.
The work of F.B. was supported by the EU grant no. ERBCHBGCT930407.

\end{document}